\documentclass[aps, prx,twocolumn,longbibliography,superscriptaddress,amsmath,amssymb,floatfix]{revtex4}
\usepackage[latin9]{inputenc}
\usepackage{amssymb}
\usepackage{graphicx}
\usepackage{amsmath}
\usepackage{color}
\usepackage{mathrsfs}
\usepackage{float}
\usepackage{indentfirst}
\usepackage{mathrsfs}
\usepackage{float}
\usepackage{indentfirst}
\usepackage{textcomp}
\usepackage{comment}
\usepackage{mathtools}
\usepackage{natbib,hyperref}
\usepackage{soul}

\begin{document}

\title{Effect of hole doping on the 120 degree order in the triangular lattice Hubbard model: A Hartree-Fock revisit}

\author{Mingpu Qin} \thanks{qinmingpu@sjtu.edu.cn}
\affiliation{Key Laboratory of Artificial Structures and Quantum Control,  School of Physics and Astronomy, Shanghai Jiao Tong University, Shanghai 200240, China}

\begin{abstract}
   We revisit the unrestricted Hartree Fock study on the evolution of the ground state of the Hubbard model on the triangular lattice with hole doping. 
   At half-filling, it is known that the ground state of the Hubbard model on triangular lattice develops a 120 degree coplanar order at half-filling in the strong
   interaction limit, i.e., in the spin $1/2$ anti-ferromagnetic Heisenberg model on the triangular lattice.
   The ground state property in the doped case is still in controversy even though extensive studies were performed in the past. 
   Within Hartree Fock theory, we find that the 120 degree order persists
   from zero doping to about $0.3$ hole doping. At $1/3$ hole doping, a three-sublattice collinear order emerges in which the doped hole is concentrated on one of
   the three sublattices with antiferromagnetic Neel order on the remaining two sublattices, which forms a honeycomb lattice. Between the 120 degree order
   and $1/3$ doping region, a phase separation occurs in which the 120 degree order coexists with the collinear anti-ferromagnetic order in different region of the system. 
   The collinear phase extends from $1/3$ doping to about $0.41$ doping, beyond which the ground state is paramagnetic with uniform electron density.
   The phase diagram from Hartree Fock could provide guidance for the future study of the doped Hubbard model on triangular lattice with more
   sophisticated many-body approaches. 
\end{abstract}

\maketitle

\section{Introduction} 
The physics of doping anti-ferromagnetic Mott insulator \cite{RevModPhys.70.1039} is one of the most important
themes in the study of correlation effect in quantum many body systems. In a Mott insulator where
charge degree of freedom is frozen, states with different types of spin order and spin liquid state can
be realized. By introducing holes into the Mott insulator, the
charge and spin degrees of freedom are entangled with each other which could result in exotic phases.
The most well known example is the emergence of high-Tc superconductivity in the cuprates by doping
the parent anti-ferromagnetic Mott insulator \cite{RevModPhys.78.17}.

The Hubbard model \cite{Hubbard238,2021arXiv210400064Q,2021arXiv210312097A} and its decedents are the simplest models to explore the physics of the doped Mott insulator.
On the square lattice, it is now known that the ground state is an antiferromagntic insulator at half-filling with even infinitesimal U value \cite{PhysRevB.94.085140}.
With doping, it was expected that superconductivity could emerge similar as in the cuprates. However, a recent collaboration work with a variety of
state-of-the-art numerical approaches established the ground 
state as stripe phase \cite{Zheng1155} with the absence of superconductivity in the pure Hubbard model \cite{PhysRevX.10.031016},
which underlines the necessity to introduce extra interacting terms in the Hubbard model to be able to describe the superconductivity in cuprates.
On the honeycomb lattice, there exists a critical interaction strength $U_c \approx 3.8$ at half filling which separates the paramagnetic
semi-metal phase in small $U$ limit and the anti-ferromagnetic insulator phase at the large $U$ limit \cite{PhysRevX.3.031010,PhysRevX.6.011029}. By doping the large
$U$ anti-ferromagnetic state, it was recently found that stripe phase emerges \cite{PhysRevB.103.155110} and superconductivity is also absent \cite{2021arXiv210414160Q},
similar as the square lattice case.    

Given that stripe order emerges by doping the antiferromagnetic Mott insulating phase on both the square and honeycomb lattices,
it is natural to ask what is the ground state of the doped Hubbard model on the triangular lattice.
Different from the bipartite square and honeycomb lattices, the triangular lattice is tripartite and hence the spin degree of freedom is frustrated on it.
For the half-filled Hubbard model in the large $U$ limit, a 120 degree coplanar order instead of a spin liquid
phase was established on the spin $1/2$ Heisenberg model on triangular lattice \cite{PhysRevLett.82.3899,PhysRevLett.99.127004}.
It is now known that the critical $U$ for the onset of the 120 degree order is $U_c \approx 12$ \cite{PhysRevB.74.085117}. But how does
the ground state evolve from small $U$ to large $U$ is still unknown. Whether the small $U$ paramagnetic metallic phase is directly connected with the 120 degree Mott insulator
phase is still in controversy \cite{PhysRevLett.100.136402,PhysRevLett.103.036401,PhysRevB.96.205130,PhysRevResearch.2.013295}. One difficulty to resolve this issue is
that the frustration in
the triangular lattice
causes the infamous sign problem in quantum Monte Carlo simulation \cite{PhysRevB.41.9301,PhysRevLett.94.170201}. Recently, possible intermediate spin liquid
phase was found by studying triangular cylinders with DMRG \cite{PhysRevB.96.205130,PhysRevX.10.021042} sparks the interest in this model \cite{2021arXiv210212904W,PhysRevB.102.115150}
with the hope that doping this spin liquid phase
could result in unconventional superconductivity \cite{2020arXiv200711963Z,2021arXiv210307998P,PhysRevB.103.165138}.

Experimentally, the discovery of superconductivity on the triangular material Na$_x$CoO$_2$.H$_2$O \cite{Takada2003} back to 2003 induced a wave of interest in the
study of Hubbard model on the triangular lattice \cite{PhysRevB.69.092504,Watanabe-2005,PhysRevLett.96.227001,PhysRevB.90.165135}.
The Hubbard model on triangular lattice model is also relevant to the recently synthesized organic compound \cite{PhysRevLett.112.177201} and  perovskite \cite{PhysRevB.95.060412}.

In this work, we will focus on the large $U$ phase where the spin is ordered 120 degree at half-filling.
We intend to investigate what is the ground state when holes are doped into the 120 degree ordered insulator.
We notice that the stripe phase found in the square \cite{Zheng1155} and honeycomb lattice \cite{PhysRevB.103.155110} can be both
obtained in Hartree-Fock calculation with renormalized interaction strength \cite{PhysRevB.40.7391,doi:10.1143/JPSJ.59.1047,PhysRevB.39.9749,Xu_2011}, we will employ the
unrestricted Hartree-Fock in this work. Though Hartree-Fock can't capture the full correlation
in the Hubbard model, the phase diagram from Hartree-Fock can provide useful guidance for
future study of the same model with more sophisticated many-body approaches \cite{PhysRevX.5.041041}. 

Hartree Fock calculation of the Hubbard model on the triangular lattice was already performed by many authors in the literature \cite{PhysRevLett.64.950,Jayaprakash_1991}. A full
Hartree Fock phase diagram can
be found in \cite{https://doi.org/10.1002/andp.19955070405}. However, in \cite{https://doi.org/10.1002/andp.19955070405}, charge degree of freedom is frozen and only
the helix spin
order was considered. In this study we don't impose any restriction on the possible spin order and also allow charge fluctuation in the Hartree-Fock solution. We find a new phase at $1/3$
doping where doped holes are concentrated on one of the three sublattices and the spin order in the other two sublattices is collinear.
Moreover, a phase separation between the 120 degree order and the $1/3$ collinear order region was found in the real space calculations.

The rest of the paper is organized as follows. In Sec.~\ref{sec_model}, we introduce the Hubbard model on triangular
lattice and the unrestricted Hartree-Fock theory. In Sec.~\ref{sec_half}, we discuss the phase diagram at half-filling.
In Sec.~\ref{sec_dope} we study the ground state away from half-filling. In Sec.~\ref{sec_phase}, we show the phase diagram.
We conclude this work with a summary and perspective
in Sec.~\ref{sec_sum}.

\begin{figure}[t]
	\includegraphics[width=80mm]{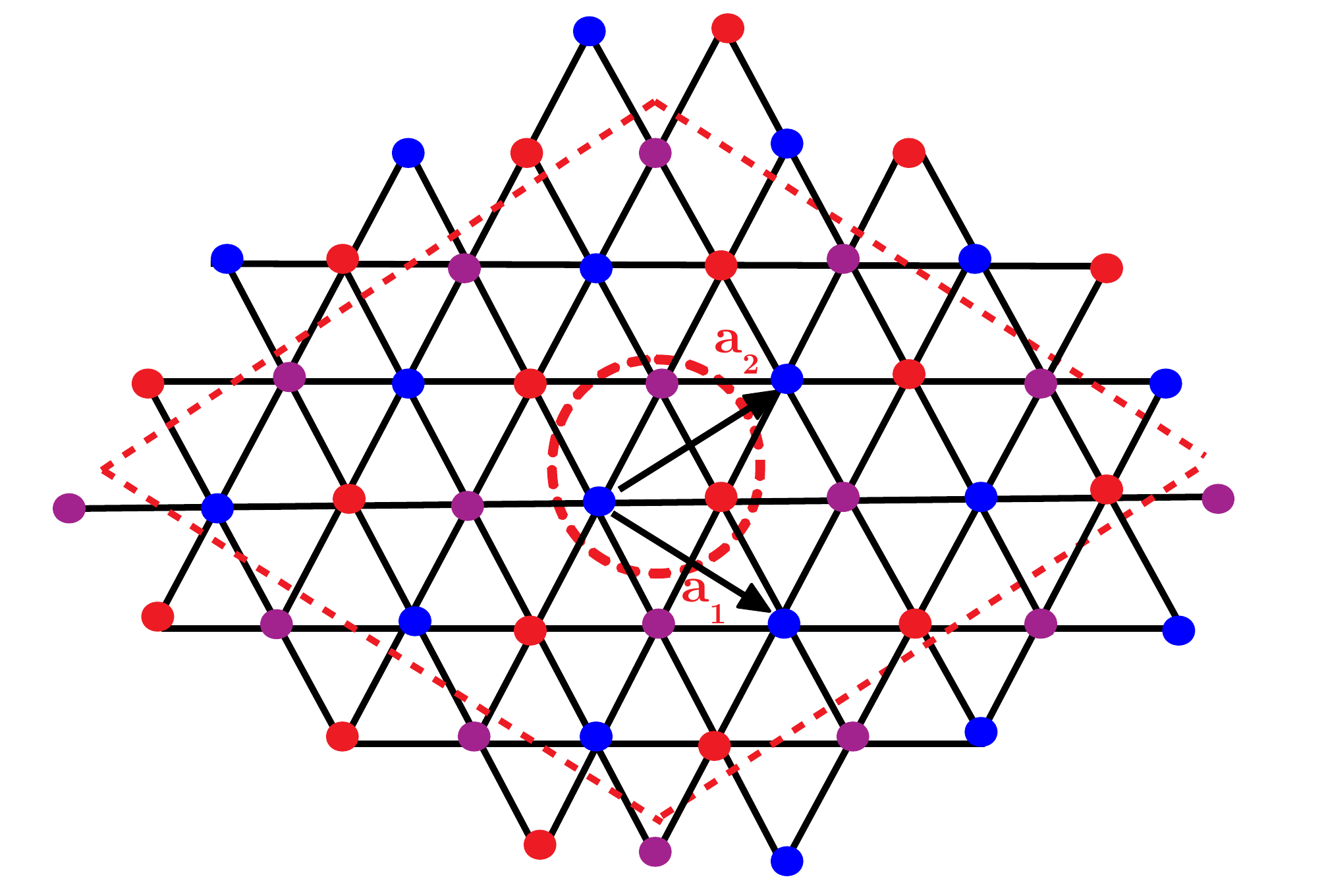}
	\caption{An illustration of the triangular lattice. We choose a three-site unit cell as shown in the dashed circle to accommodate possible
		orders with three sublattices. The two primitive vectors
		are $a_1$ and $a_2$. In the dotted rhombus, a supercell with $L_1 \times  L_2 = 3 \times 3$ is shown.
	Periodic boundary conditions are adopted in all calculations.} 
	\label{lattice}
\end{figure} 

\section{Model and Computational methods}
\label{sec_model}

\subsection{Model}
The Hamiltonian of the Hubbard model is:
\begin{equation}
	H=-t\sum_{\langle i,j\rangle,\sigma}c_{i\sigma}^\dagger c_{j\sigma}+U\sum_{i}n_{i\uparrow}n_{i\downarrow}
	\label{ham}
\end{equation}
where $t$ is the hopping constant and is set to the energy unit.
The density is denoted as $n = N_e/N_s$ with $N_e$ and $N_s$ the total number of electrons and lattice sites.
The hole doping is $h = 1 - n$.
The local spin in $z$ and $x$ direction at site $i$ are $m_{i}^z = (\langle n_{i,\uparrow} \rangle - \langle n_{i, \downarrow} \rangle) / 2$ and
$m_{i}^{x} = (c_{i\uparrow}^{\dagger}c_{i\downarrow}+c_{i\downarrow}^{\dagger}c_{i\uparrow}) / 2$ respectively.
The hole density at site $i$ is $h_i = (1 - \langle n_{i,\uparrow} \rangle - \langle n_{i, \downarrow} \rangle)$. 
We study the Hubbard model on the triangular lattice which is shown in Fig.~\ref{lattice}. We choose a three-site 
unit cell as in the dashed circle of Fig.~\ref{lattice}. The two primitive vectors are $a_1$ and $a_2$. Periodic boundary conditions are adopted in all calculations.
We only
consider nearest neighboring hopping and focus on hole doping ($h\ge0$) in this work.

    \begin{figure}[t]
	\includegraphics[width=80mm]{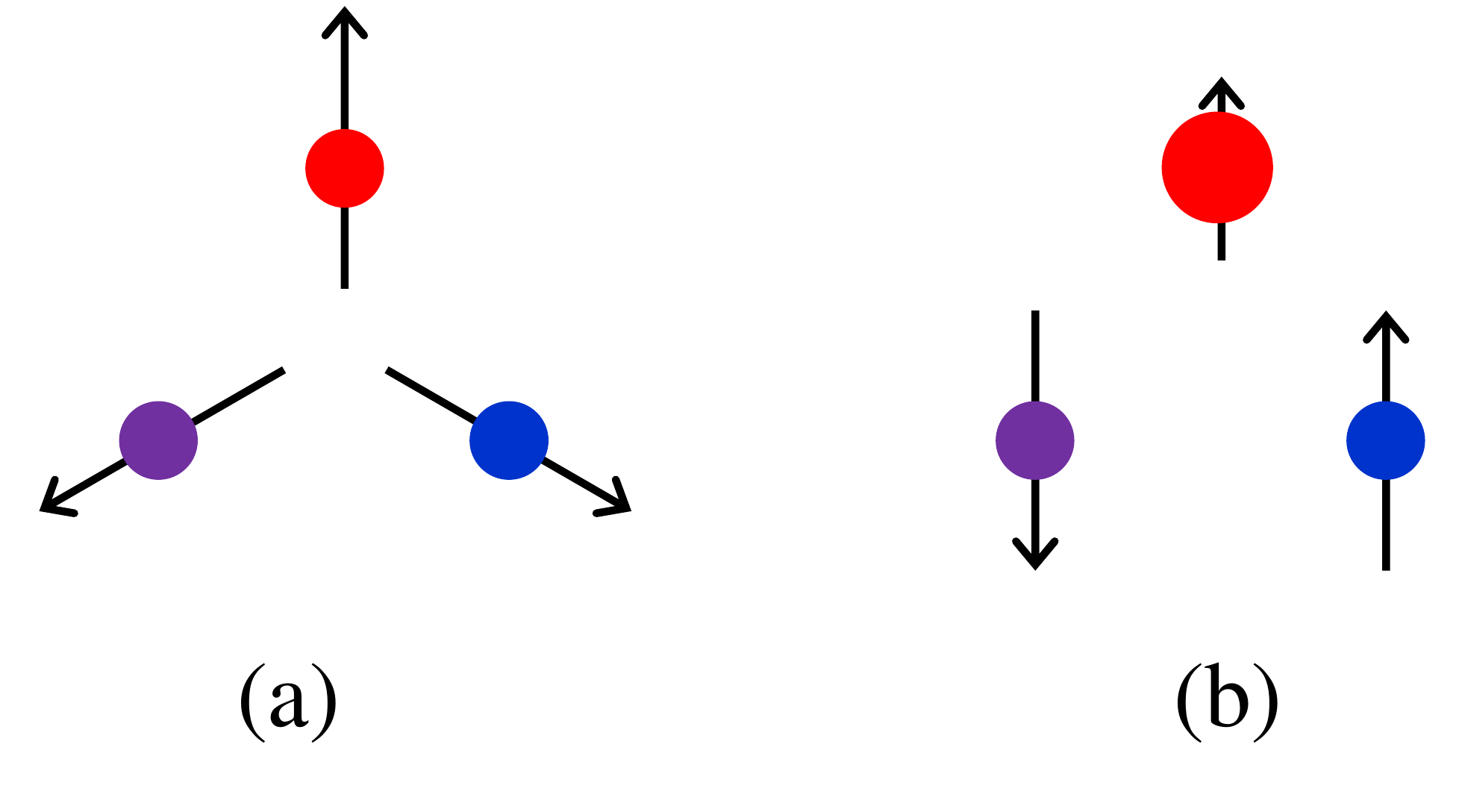}
	\caption{The two types of orders considered in the hole doped region. The arrow represents the local spin density and the area of the circle is proportional to the
		hole density on each site. (a): the 120 degree order. (b): the collinear order. In (a), the charge (hole)
		is uniformly distributed, while a charge order is also present in (b). The related strength of collinear order among the three sublattices can
		vary with doping in (b). At the special $1/3$ doping, the order is like the antiferromagnetic order on honeycomb lattice in which the spin order is exactly zero for one sublattice and
		the doped holes are concentrated on the same sublattice.} 
	\label{order}
\end{figure} 

\subsection{Hartree-Fock theory}

In the unrestricted Hartree-Fock theory, the Hamiltonian in Eq.~(\ref{ham}) is decoupled as

\begin{equation}
\begin{split}
H_{hf} = &-t\sum_{\langle ij\rangle\sigma}c_{i\sigma}^{\dagger}c_{j\sigma}+U\sum_{i}(\langle n_{i\uparrow}\rangle n_{i\downarrow}
+\langle n_{i\downarrow}\rangle n_{i\uparrow} \\
&-\langle S_{i}^{+}\rangle S_{i}^{-} -\langle S_{i}^{-}\rangle S_{i}^{+}-\langle n_{i\uparrow}\rangle\langle n_{i\downarrow}\rangle+\langle S_{i}^{+}\rangle\langle S_{i}^{-}\rangle) 
\label{ham_hf}
\end{split}
\end{equation}
where $n_{i\uparrow} =c_{i\uparrow}^{\dagger}c_{i\uparrow}$, $n_{i\downarrow} =c_{i\downarrow}^{\dagger}c_{i\downarrow}$,
$S_{i}^{+} =c_{i\uparrow}^{\dagger}c_{i\downarrow}$, and $S_{i}^{-} =c_{i\downarrow}^{\dagger}c_{i\uparrow}$.
To obtain the Hartree Fock solution, we need to solve Eq.~(\ref{ham_hf}) self-consistently.
The strategy we adopt is as follows. We first solve Eq.~(\ref{ham_hf}) self-consistently in real space with different sizes to search for the possible orders.
In the real space calculation, we don't impose any restriction on the direction of possible spin order. 
The procedure is described as follows. We first choose an initial set of values for $\{ \langle n_{i\uparrow}\rangle, \langle n_{i\downarrow}\rangle, \langle S_{i}^{+}\rangle, \langle S_{i}^{-}\rangle \} $, and
	feed it into Eq.~(\ref{ham_hf}) which is then a free Hamiltonian and can be easily solved. From the solution of Eq.~(\ref{ham_hf}) we can calculate the new values for
	$\{ \langle n_{i\uparrow}\rangle, \langle n_{i\downarrow}\rangle, \langle S_{i}^{+}\rangle, \langle S_{i}^{-}\rangle \} $. This process is repeated till
	convergence. In the real space calculations, the largest system size we used is $24 \times 24$ which corresponds to $1728$ sites. We try different
	initial configurations and find two possible orders, i.e, the 120 degree order near half-filling and a collinear order near $1/3$ doping in the large $U$
	region (see Fig.~\ref{order}). We also find a phase separation region in between. The details of the results are discussed in the following sections.

Then we turn to momentum space with
unit cell compatible to the real space solution which allows us to study systems with larger size. We also compare the energies
of different types of orders to determine which one is the true ground state. With the unit cell chosen in Fig.~\ref{lattice}, the
Hartree-Fock Hamiltonian in $k$ space is (from a Fourier transformation of Eq.~(\ref{ham_hf}))
\begin{widetext}
\begin{equation}
H=-t\sum_{k}\left(\begin{array}{cccccc}
c_{k\uparrow}^{\dagger A} & c_{k\uparrow}^{\dagger B} & c_{k\uparrow}^{\dagger C} & c_{k\downarrow}^{\dagger A} & c_{k\downarrow}^{\dagger B} & c_{k\downarrow}^{\dagger C}\end{array}\right)\left[\begin{array}{cccccc}
\frac{n_{A}-2m_{A}^{z}}{2}U & A(k) & A^{*}(k) & -m_{A}^{x}U & 0 & 0\\
A^{*}(k) & \frac{n_{B}-2m_{B}^{z}}{2}U & A(k) & 0 & -m_{B}^{x}U & 0\\
A(k) & A^{*}(k) & \frac{n_{C}-2m_{C}^{z}}{2}U & 0 & 0 & -m_{C}^{x}U\\
-m_{A}^{x}U & 0 & 0 & \frac{n_{A}+2m_{A}^{z}}{2}U & A(k) & A^{*}(k)\\
0 & -m_{B}^{x}U & 0 & A^{*}(k) & \frac{n_{B}+2m_{B}^{z}}{2}U & A(k)\\
0 & 0 & -m_{C}^{x}U & A(k) & A^{*}(k) & \frac{n_{C}+2m_{c}^{z}}{2}U
\end{array}\right]\left(\begin{array}{c}
c_{k\uparrow}^{A}\\
c_{k\uparrow}^{B}\\
c_{k\uparrow}^{C}\\
c_{k\downarrow}^{A}\\
c_{k\downarrow}^{B}\\
c_{k\downarrow}^{C}
\end{array}\right)
\end{equation}
\label{hf_mom}
\end{widetext}
with $A(k)=e^{ik\delta_{1}}+e^{-ik\delta_{2}}+e^{ik(\delta_{2}-\delta_{1})}$ and $\delta_{1} =(\frac{3}{2},-\frac{\sqrt{3}}{2}),\delta_{2}=(\frac{3}{2},\frac{\sqrt{3}}{2})$. For
the 120 degree order, we set $n_A = n_B = n_C = n/3$, $m_A^z = m, m_A^x = 0$, $m_B^z = m/2, m_B^x = -\sqrt{3}m/2$, $m_C^z = m/2, m_C^x = \sqrt{3}m/2$. For the collinear
order (see Fig.~\ref{order}),
we set $n_A + n_B + n_C = n$ and $m_A^x = m_B^x = m_C^x = 0$ to restrict the spin of all three sublattices in $z$ direction, but allow the
magnitude of spin order and the hole density to fluctuate freely.

\section{Half-filling}
\label{sec_half}
\begin{figure}[t]
	\includegraphics[width=80mm]{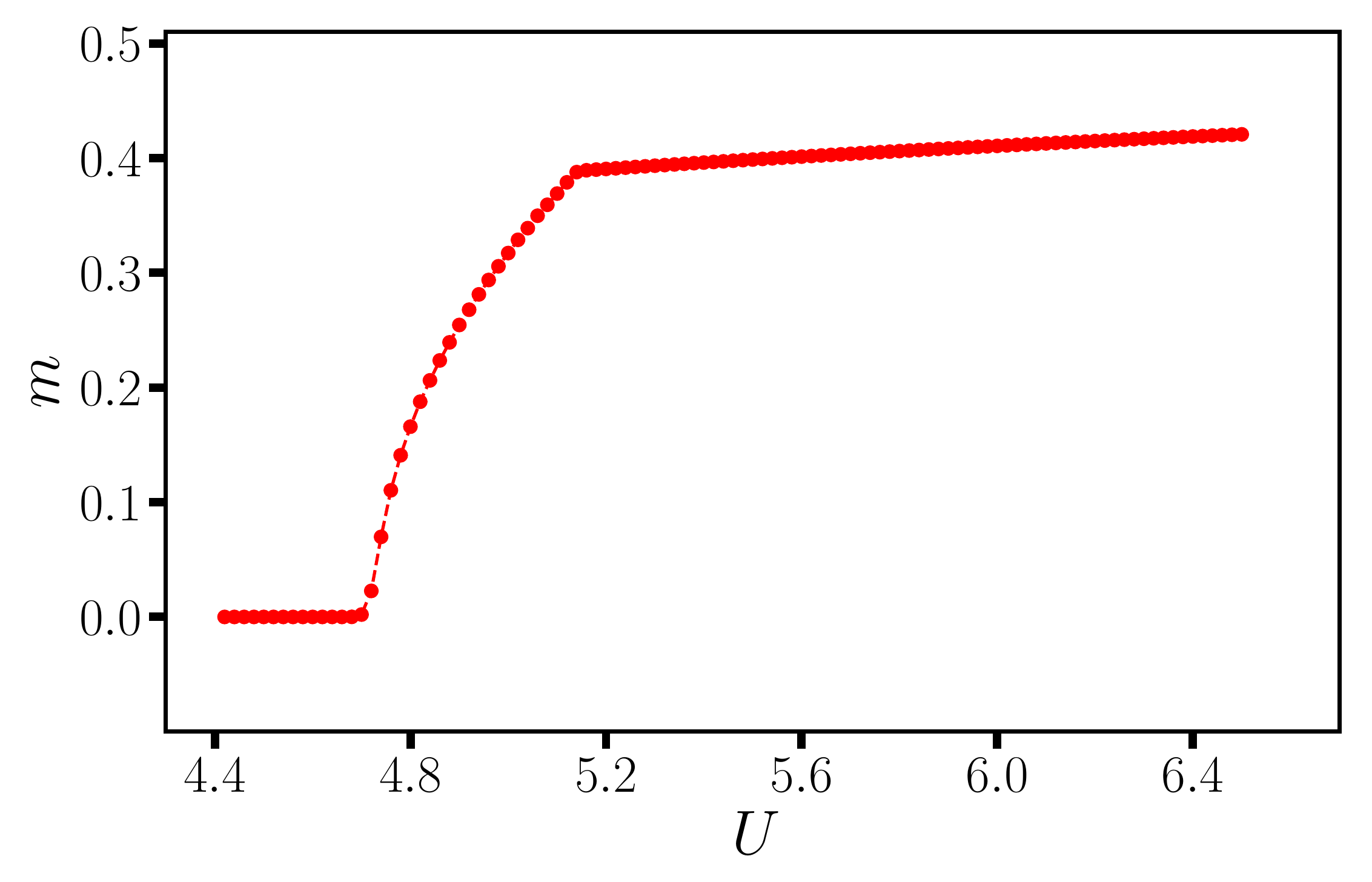}
	\includegraphics[width=80mm]{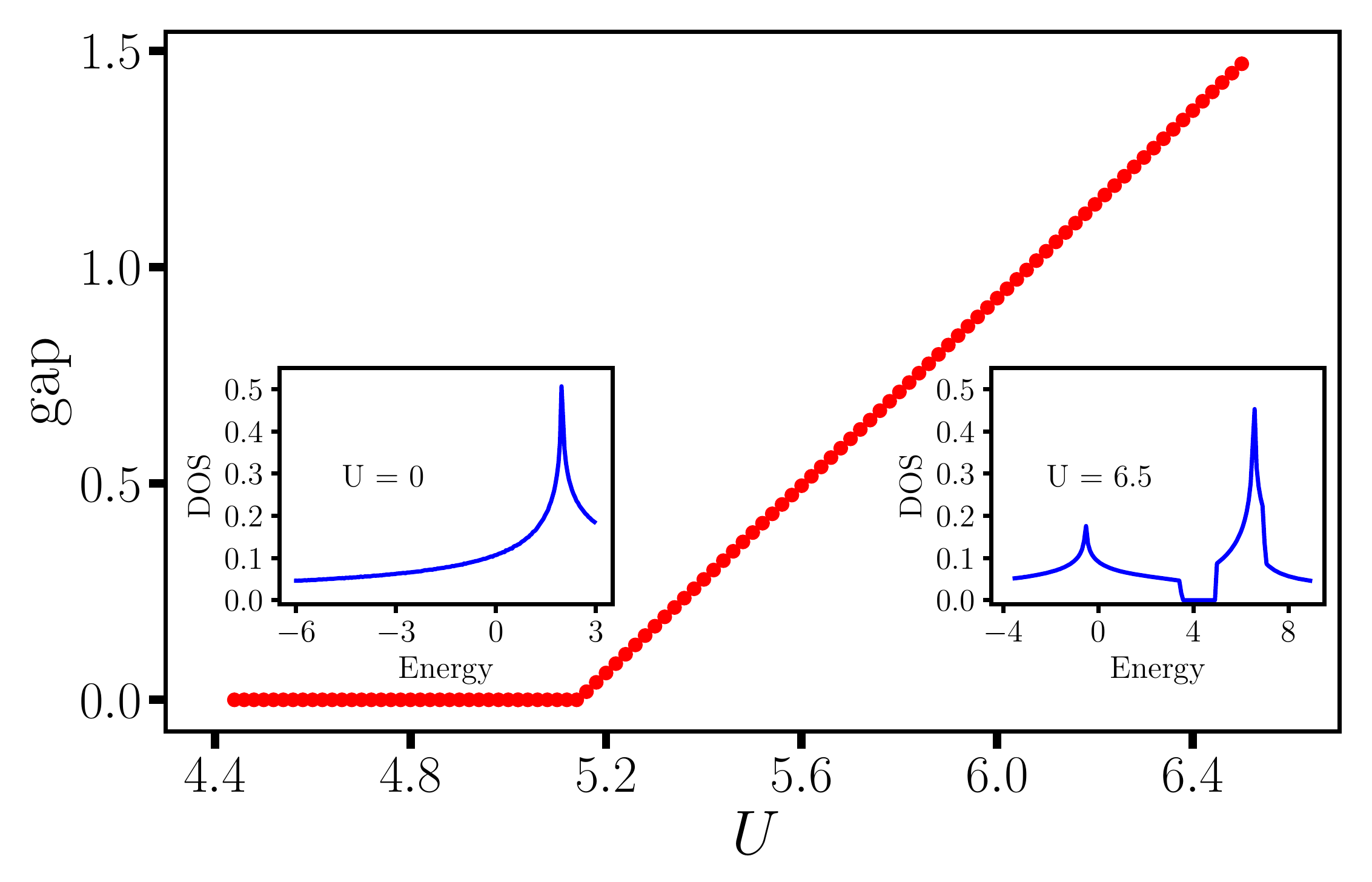}
	\caption{Upper: magnetization of the 120 degree order at half-filling versus $U$. We can find two phase transitions
		at half filling with $U_{c1} \approx 4.7$ and $U_{c2} \approx 5.14$. 
		Lower: the evolution of charge gap with $U$. We can see that charge gap only opens for $U > U_{c2}$. 
		In the inset of the lower panel, the density of state is plotted for $U = 0$ and $U = 6.5$. We need to mention that between $U_{c1}$ and $U_{c2}$,
	the helix order with varying wave vector has lower energy \cite{https://doi.org/10.1002/andp.19955070405}, which means the 120 degree order develops only above $U_{c2}$.
     We calculate a $300 \times 300$ system in the momentum space (please notice that a three-site unit cell is chosen in this work as shown in Fig.~\ref{lattice})
     	to get rid of the finite size effect. } 
	\label{mag_half}
\end{figure}

It is known that the half-filled Hubbard model on the triangular lattice is a Mott insulator with 120 degree spin order in the
large $U$ limit \cite{PhysRevLett.82.3899,PhysRevLett.99.127004}.
In Fig.~\ref{mag_half}, we show the evolution of magnetization of the 120 degree order and the charge gap with $U$ for a $300 \times 300$ lattice.
By fixing the order type and the unit cell, we can carry out the calculation in the momentum space with very large sizes. As can be seen in Fig.~\ref{mag_half}, there
are two phase transitions at half-filling by varying $U$ values. At $U \approx 5.14$ we can find
a first order transition. There is
another phase transition at $U \approx 4.7$. 
However, the charge gap only opens for $U > U_{c2}$. It was known that between $U_{c1}$ and $U_{c2}$ the true ground state is a helix order with the
wavevector varying with
$U$ \cite{https://doi.org/10.1002/andp.19955070405}. So the 120 degree order only develops for $U > U_{c2}$.  Hartree-Fock usually overestimates the magnitude of the order so the critical $U_{c2}$ here is smaller than the value from
more accurate many-body calculations \cite{PhysRevB.74.085117}. We notice that there were studies indicating there is an intermediate spin liquid phase between the small $U$ paramagnetic 
and the large $U$ 120 degree order \cite{PhysRevB.96.205130,PhysRevX.10.021042} region. It will be interesting to compare the intermediate helix spin phase
 \cite{https://doi.org/10.1002/andp.19955070405} with the spin liquid phase. We also notice that
 an intermediate magnetically disordered phase was also obtained for the 
 $\pi$ flux Hubbard model on triangular lattice \cite{PhysRevLett.114.167201}.

\begin{figure*}[t]
	\includegraphics[width=170mm]{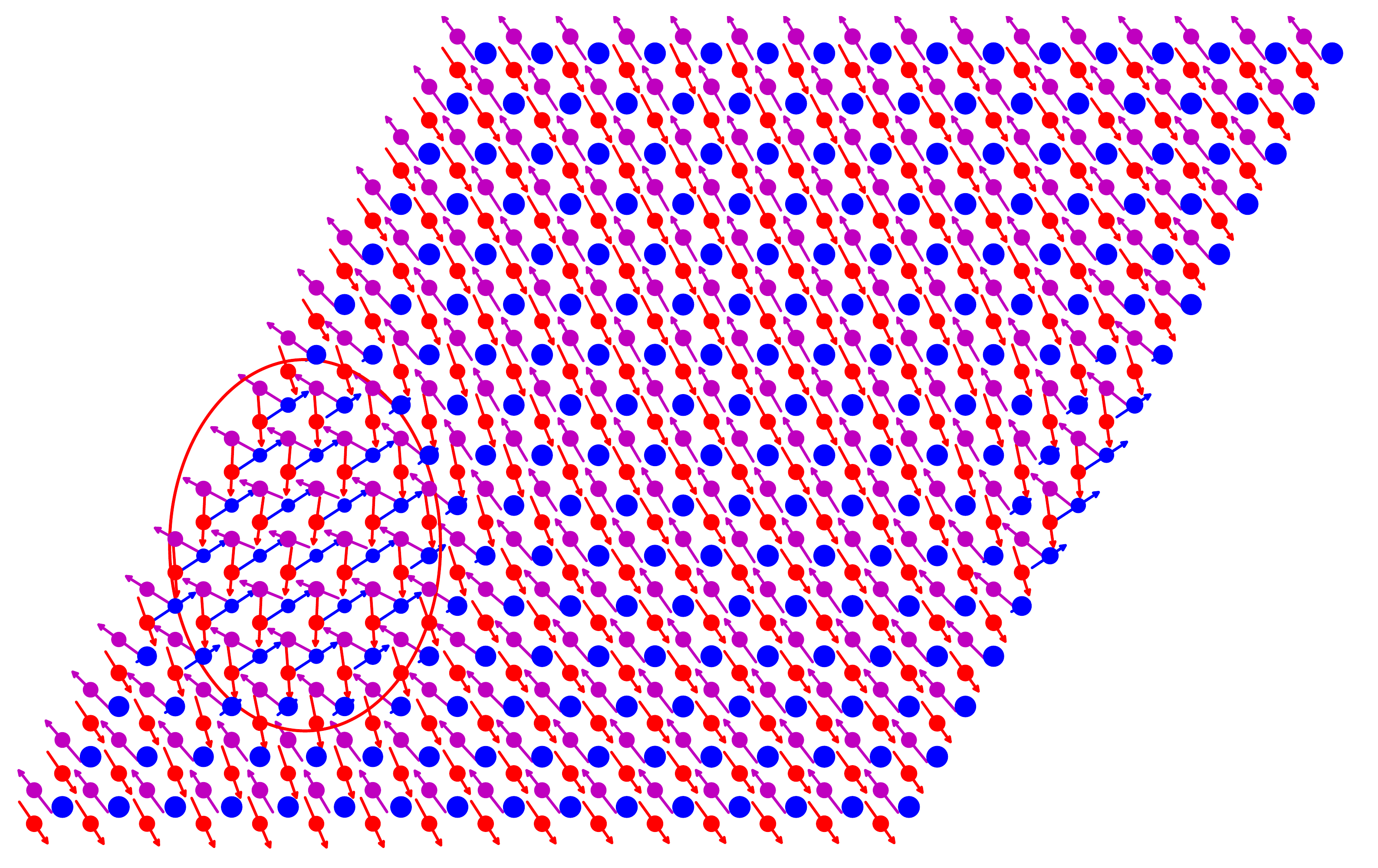}
	\caption{Spin and hole density for a $16 \times 16$ lattice. The calculation is performed in the real space by solving Eq.~(\ref{ham_hf})
		self-consistently. The arrow represents the local spin density and the area
		of the circle is proportional to the hole density on each site. 8 extra electrons are introduced into the $1/3$ doped collinear order.
		We can clearly find a phase separation in the system. The collinear order is present in most part of the system, while the 120 degree
		order can be found in the small region marked by the red oval. Similar result is also obtained for a larger $24 \times 24$ system.} 
	\label{phase_sep}
\end{figure*}

\section{Away from half-filling}
\label{sec_dope}
Doping an antiferromagnetic Mott insulator usually results in exotic phases because of the competition of kinetic and potential energies.
In this section and hereafter, we will study how the 120 degree order evolves with hole doping by fixing $U = 6$ which is larger than $U_{c2}$,
to ensure the parent state is in the 120 degree phase. 

\subsection{The stability of the 120 degree order}
We first study the small doping region to study the stability of the 120 degree order.
we find that the 120 degree order is quite robust against hole doping. It persists to doping as large as $h = 0.3$ with $U = 6$.
This result is quite different from the case of the square and honeycomb lattices, where the antiferromagnetic Neel order
turns to stripe order with small doping \cite{Zheng1155,PhysRevB.103.155110}. The coordination number on
triangular lattice ($6$) is larger than that of square ($4$) and honeycomb lattices ($3$). So to
develop a stripe like order, the cost of interacting energy will be higher, which might not be able to
fully compensated by the gain of kinetic energy in a stripe state.

\subsection{Collinear order at $1/3$ doping}
As expected, the 120 degree order finally melts with larger doping. By increasing the doping to $1/3$, we observe a 
collinear phase with
both spin and charge order as in Fig.~\ref{order} (b). At exact $1/3$ doping, the spin on one of the three sublattices is absolutely zero and
the holes are concentrated on the same sublattice. The spin on the other two sublattices is like the antiferromagnetic order on the honeycomb lattice.
In the infinite U limit, all the doped hole resides on one sublattice and the triangular lattice is reduced to the honeycomb lattice.
We notice that this order was previously observed in different calculations \cite{doi:10.1143/JPSJ.74.2901,PhysRevB.70.020504,PhysRevLett.113.246405}.

\begin{figure}[t]
	\includegraphics[width=80mm]{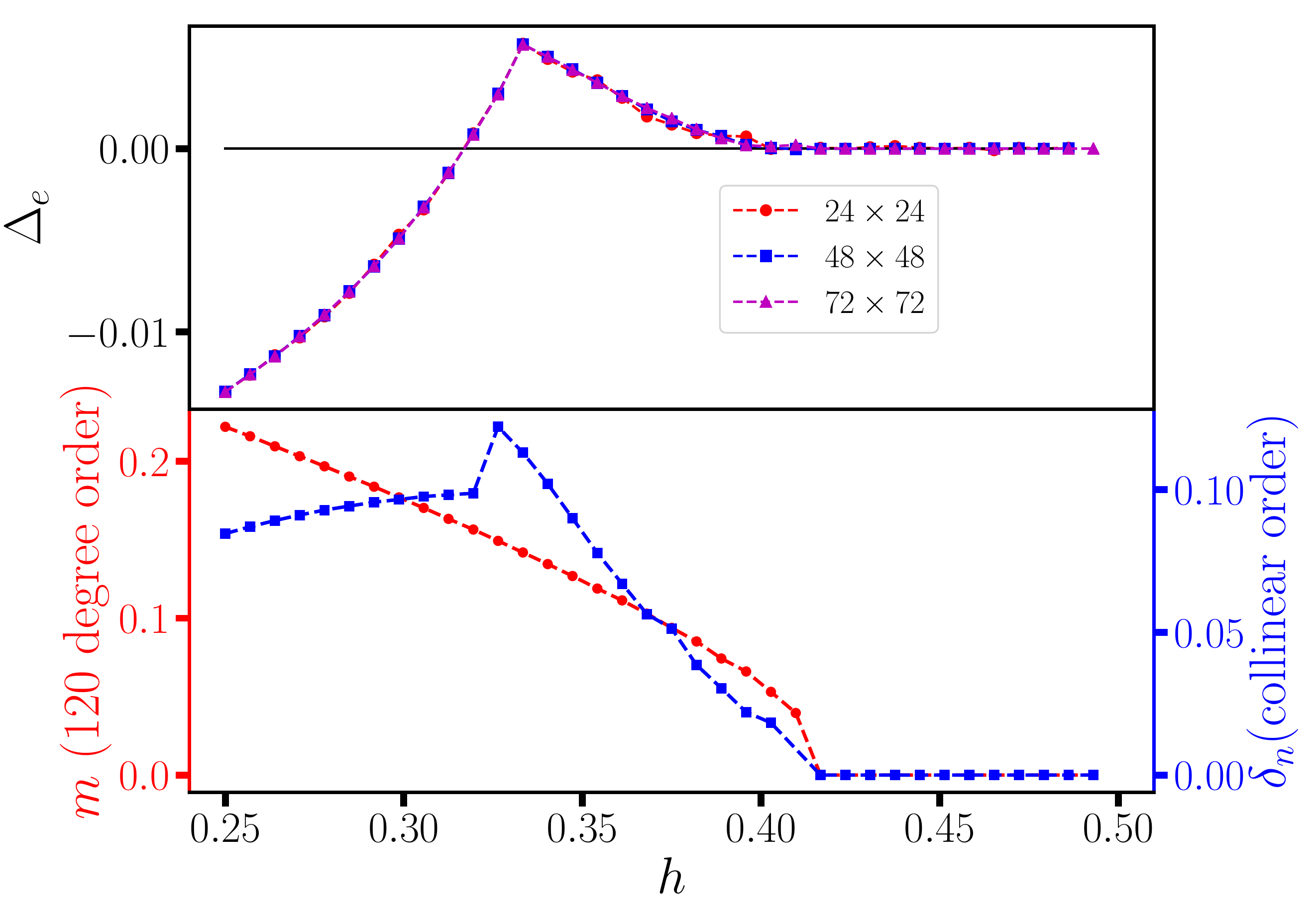}
	\caption{Upper: energy difference between the 120 degree order and the collinear order.	
		Lower: the order parameters for both the 120 degree order and the collinear order. For the 120 degree order,
		the order parameter is the magnetization, while the collinear order parameter is defined as the largest difference of densities
		among three sub-lattices. Please
		note that the results of order parameter are obtained by
		restricting the solution to a specific order when solving
		the Hartree-Fock equation. So a finite order parameter
		doesn't necessarily mean the ground state has the corresponding order. We need to compare the energies (as in the upper panel)
		to determine the order in the real ground state.	
	Between $0.3$ and $1/3$ doping, a phase separation occurs (see the discussion in Sec.~\ref{phase_sep}). For doping
    larger than $0.41$, the order parameters for both 120 degree and the collinear order are zero,
    which indicates the ground state is paramagnetic with uniform charge density in this region.
    } 
	\label{delta_e}
\end{figure}

In the vicinity of the $1/3$ doping, the spin order remains collinear with the magnitude of the spin and hole order changes with
doping. In the upper panel of Fig.~\ref{delta_e}, we plot the energy difference between the 120 degree order and the collinear order. 
	The evolution of the order parameters for both the 120 degree order and the collinear order with doping is shown in the lower panel.
	Please note that the results of order parameter are obtained by restricting the solution to a specific order when solving the Hartree-Fock equation.
	So a finite order parameter doesn't necessarily mean the ground state has the corresponding order. We need to compare the energies (as in the upper panel of Fig.~\ref{delta_e})
	to determine the order in the real ground state.       
Finite size effect of the energy difference is small as shown by the results from
size of $24 \times 24$ to $72 \times 72$. The peak position of the energy difference in Fig.~\ref{delta_e} is exactly at $1/3$ doping.
As we will discuss below, between $0.3$ and $1/3$ doping, a phase separation \cite{PhysRevLett.64.475} occurs where the 120 degree coexists with the colliear order
in different region of the system. For doping larger than $1/3$, the
collinear phase has lower energy. When doping is larger than $0.41$, the system is paramagnetic with uniform charge density where the orders are
zero and the two types of states have equal energy (can be seen from the lower panel of Fig.~\ref{delta_e}). We notice that this collinear phase can be
viewed as a stripe phase with periodicity $3$ for both spin and
charge order, even though it is totally different from the parent 120 degree order.   

\subsection{The phase separation}
Between $h = 0.3$ and $ h = 1/3$, we find a phenomenon of phase separation \cite{PhysRevLett.64.475} in the real space calculation. In Fig.~\ref{phase_sep},
we study a $16 \times 16$ lattice by adding 8 extra electrons to the $1/3$ doped system. We can clearly find a phase separation
from the spin and hole density distribution in Fig.~\ref{phase_sep}. Most part of the system displays an order same as the
$1/3$ doped collinear order, while a 120 degree order emerges in a small portion of the system as indicated by the oval of Fig.~\ref{phase_sep}.
The position of the boundary separating the two regions depends on the initial configuration used in the self-consistent calculation. 
The occurrence of phase separation in this region means a uniform state has higher energy \cite{PhysRevLett.64.475}.
Near the boundary of the two phases, we find that the spin order was distorted to connect the two phase smoothly in order not to cause a large increase of
energy. 
With the decrease of doping, the area of 120 degree order region expands and the system eventually switches to the 120 degree order
with $h < 0.3$. On the square lattice, phase separation in Hubbard model was usually predicted in the vicinity of half-filling \cite{PhysRevB.94.195126}. 
But here on the triangular lattice, it occurs in the boundary of 120 degree order phase and the collinear phase.

\section{The phase diagram}
\label{sec_phase}

\begin{figure}[t]
	\includegraphics[width=80mm]{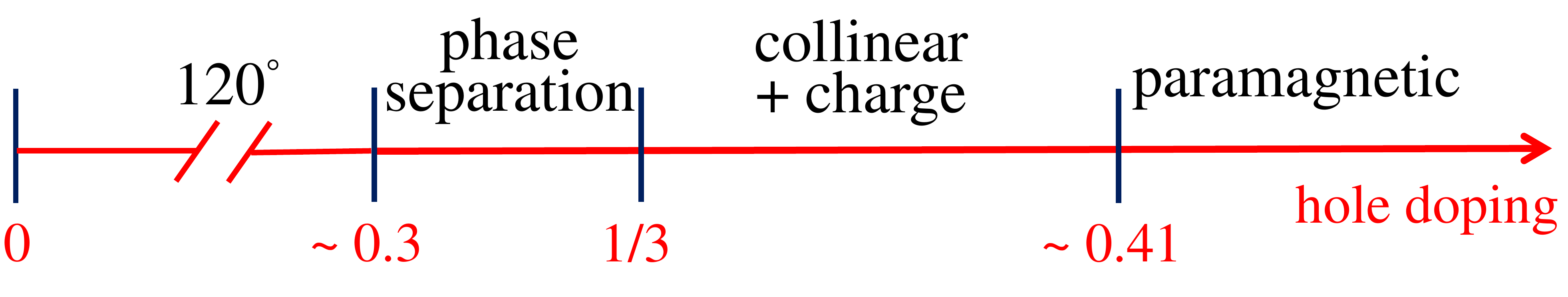}
	\caption{The Hatree Fock phase diagram of the hole doped Hubbard model on triangular lattice with $U = 6$.} 
	\label{phase_diag}
\end{figure}

The whole phase diagram with $U = 6$ is summarized in Fig.~\ref{phase_diag}. The 120 degree order extends from half-filling to
about $0.3$ doping. Between $1/3$ and $0.4$ doping, collinear phase with charge order is the ground state with a phase 
separation region connecting the 120 degree order and the collinear phase. With doping larger than $0.4$, the
ground state is paramagnetic with uniform charge density.

%

\section{Summary and perspectives}
\label{sec_sum}
We revisit the Hartree-Fock calculation of the evolution of the 120 degree order of the Hubbard model on the triangular lattice with hole doping.
We find that the 120 degree order is quite stable against hole doping and persists to hole doping as large as $0.3$. We find a collinear phase
in the vicinity of $1/3$ doping. This collinear order is similar to the antiferromagnetic order on the honeycomb lattice with doped hole concentrated on one of the
three sublattices. Between $0.3$ and $1/3$ doping, a phase separation occurs where the 120 degree order coexists with the collinear order in different
region of the system.

Different from the square and honeycomb lattices, doped hole doesn't cause the antiferromagnetic order turns to a stripe order
in the small doping region. However, the collinear phase near $1/3$ doping can be viewed as a stripe phase with
the periodicity of spin and charge order $\lambda = 3$, even though the collinear order is totally different from
the parent 120 degree order.

The collinear order in the vicinity of $1/3$ doping was previously observed in the $1/3$ electron doped region with the slave-boson approach \cite{PhysRevB.90.165135},
in the t-J-V model with series expansion and cluster mean-field theory \cite{PhysRevB.70.020504}, and in the extended Hubbard model with variational Monte
Carlo method \cite{PhysRevLett.113.246405,doi:10.1143/JPSJ.74.2901}. The final determination of the existence of the collinear order in the pure Hubbard model on
triangular lattice need
calculations with accurate many-body methods.

Superconductivity was predicted in the Hubbard model on triangular lattice by doping the intermediate spin liquid phase \cite{2021arXiv210307998P,PhysRevB.103.165138}.
In this work we focus on the large $U$ region to study the doping effect on the 120 degree order and
the study of superconductivity is beyond the reach of unrestricted Hartree Fock in which particle number is conserved.
In \cite{2020arXiv200711963Z}, it was found that pairing correlation is enhanced when doping $h > 20\%$ in the large $U$ region.
Again, the possibility of collinear order in this region when many body correlation is considered needs further investigation. 

We notice a recent Hartree-Fock calculation \cite{PhysRevB.104.075150} on the same model but with a phase on the hopping term which
is relevant to twisted homobilayer WSe$_2$.



\begin{acknowledgments}
This work is supported by a start-up fund from School of Physics and Astronomy in Shanghai Jiao Tong University. 
We thank useful discussions with T. Xiang.
\end{acknowledgments}

\bibliography{triangular.bib}

\appendix

\end{document}